\documentclass[preprint,superscriptaddress]{revtex4-1}

\usepackage{graphicx}% Include figure files
\usepackage{dcolumn}% Align table columns on decimal point
\usepackage{bm}% bold mat
\usepackage{amsmath}
\usepackage{amssymb}

\usepackage{subfig}

\def\gs{\gtrsim}
\def\ls{\lesssim}
\def\fr{\frac}
\def\pa{\partial}

\def\be{\begin{equation}}
\def\ee{\end{equation}}
\def\bea{\begin{eqnarray}}
\def\eea{\end{eqnarray}}

\def\fr{\frac}
\def\pa{\partial}
\def\ti{\tilde}
\def\ep{\epsilon}
\def\vae{\varepsilon}
\def\la{\lambda}
\def\vap{\varphi}
\def\intel{\int d^4x\sqrt{-g}}
\def\bg{\bar{g}}
\def\bR{\bar{R}}
\def\bn{\bar{\nabla}}
\def\bT{\bar{T}}
\def\mpl{M_{\rm Pl}}
\def\Ve{V_{\rm eff}}

\def\pmin{\phi _{\rm min}}
\def\mn{{\mu \nu}}
\def\mO{\mathcal{O}}
\def\mH{\mathcal{H}}

\def\rhom{\rho _{\rm m}}
\def\rhode{\rho _{\rm DE}}
\def\Pn{\Phi _{\rm N}}

\def\dph{\delta \phi}
\def\dphi{\delta \phi _{\rm in}}

\def\dpho{\delta \phi _{\rm out}}
\def\eth{\ep _{\rm th}}

\begin{document}

\title{Equation of state of dark energy in $f(R)$ gravity}

\author{Kazufumi Takahashi}
	\email{Email: ktakahashi@resceu.s.u-tokyo.ac.jp}
	\affiliation{Department of Physics, Graduate School of Science,\\
	The University of Tokyo, Tokyo 113-0033, Japan}
	\affiliation{Research Center for the Early Universe (RESCEU), Graduate School of Science,\\
	The University of Tokyo, Tokyo 113-0033, Japan}
\author{Jun'ichi Yokoyama}
	\email{Email: yokoyama@resceu.s.u-tokyo.ac.jp}
	\affiliation{Research Center for the Early Universe (RESCEU), Graduate School of Science,\\
	The University of Tokyo, Tokyo 113-0033, Japan}
	\affiliation{Kavli Institute for the Physics and Mathematics of the Universe (Kavli IPMU),\\
	WPI, TODIAS, The University of Tokyo, Kashiwa, Chiba, 277-8568, Japan}

\begin{abstract}
$f(R)$ gravity is one of the simplest generalizations of general relativity, which may explain the accelerated cosmic expansion without introducing a cosmological constant. Transformed into the Einstein frame, a new scalar degree of freedom appears and it couples with matter fields. In order for $f(R)$ theories to pass the local tests of general relativity, it has been known that the chameleon mechanism with a so-called thin-shell solution must operate. If the thin-shell constraint is applied to a cosmological situation, it has been claimed that the equation-of-state parameter of dark energy $w$ must be extremely close to $-1$. We argue this is due to the incorrect use of the Poisson equation which is valid only in the static case. By solving the correct Klein-Gordon equation perturbatively, we show that a thin-shell solution exists even if $w$ deviates appreciably from $-1$.
\end{abstract}

\maketitle

%%%%%%%%%%%%%%%%%%%%%%%%%%%%%%%%%%%%%%%%%%%%%%%%%%%%%%%%%%%%%%%%%%%%%%%%%%%%%%%%%%%%
%%%%%%%%%%%%%%%%%%%%%%%%%%%%%%%%%%%%%%%%%%%%%%%%%%%%%%%%%%%%%%%%%%%%%%%%%%%%%%%%%%%%
%	Introduction
%%%%%%%%%%%%%%%%%%%%%%%%%%%%%%%%%%%%%%%%%%%%%%%%%%%%%%%%%%%%%%%%%%%%%%%%%%%%%%%%%%%%
%%%%%%%%%%%%%%%%%%%%%%%%%%%%%%%%%%%%%%%%%%%%%%%%%%%%%%%%%%%%%%%%%%%%%%%%%%%%%%%%%%%%
\section{Introduction}

The cosmic acceleration was discovered from observations of type Ia supernovae \cite{Riess:1998cb,Perlmutter:1998np}. The mechanism which causes the acceleration is still not clear and many approaches have been tried. The simplest one is the $\Lambda$ cold dark matter ($\Lambda$CDM) model, which is based on Einstein's general relativity (GR) and consists of CDM and a cosmological constant $\Lambda$. In this model, however, there is a problem in that the parameter must be fine-tuned in order to explain the observed energy budget of the Universe.

Another approach is to modify the theory of gravity from GR, and many models have been proposed (for a review, see Ref.~\cite{Tsujikawa:2010zza}). One of these is a class of $f(R)$ theories \cite{Starobinsky:2007hu,Hu:2007nk,Sotiriou:2008rp,DeFelice:2010aj}, which is the easiest generalization of GR. These models can be recast into the form of GR plus a scalar field by a conformal transformation \cite{Maeda:1988ab}. In this Einstein frame, the scalar couples with a matter field and matter experiences the fifth force. This new type of force must be small in order to pass the local tests of gravity \cite{Hoskins:1985tn,Will:2005va,Talmadge:1988qz,Vessot:1980zz}. To avoid this difficulty, it is necessary that the so-called chameleon mechanism operates and a thin-shell scalar configuration exists \cite{Khoury:2003rn}. If an object has a thin shell, the mass of the scalar field gets large inside the object and the fifth force is suppressed. As an alternative model for dark energy, it is quite important to get a constraint on the effective equation-of-state parameter $w$ from the viewpoint of distinguishing models. If the thin-shell constraint is naively applied to a cosmological situation, it has been claimed that $w$ must be extremely close to $-1$ \cite{Brax:2008hh} (see also Ref.~\cite{Wang:2012kj}). However, this constraint is physically unacceptable since a cosmological background quantity like $w$ should not be constrained only by local information. In this paper, we perform a more precise analysis and show that the previous constraint of Ref.~\cite{Brax:2008hh} does not apply.

This paper is organized as follows. In Sec.~\ref{f(R)}, we review the basics of $f(R)$ theories. The thin-shell solution, which explains how GR is restored in $f(R)$ gravity in a local scale, is also described for a static case. In Sec.~\ref{EoS}, we analyze $f(R)$ cosmology and consider the constraint on $w$. Here we obtain the scalar field configuration with a thin shell in the situation where the Universe is dominated by dark energy with $w\ne -1$.

Throughout this study, we use natural units with $c=\hbar =1$. The reduced Planck mass is written as $\mpl =(8\pi G)^{-1/2}=\kappa ^{-1}$. The sign convention is as follows.
\begin{itemize}
\item The metric has signature $(-,+,+,+)$.
\item The Riemann tensor is defined as $R^\lambda {}_{\mu \nu \sigma}=\Gamma ^\lambda {}_{\mu \sigma ,\nu}-\Gamma ^\lambda {}_{\mu \nu ,\sigma}+\Gamma ^\lambda {}_{\alpha \nu}\Gamma ^\alpha {}_{\mu \sigma}-\Gamma ^\lambda {}_{\alpha \sigma}\Gamma ^\alpha {}_{\mu \nu}$.
\item The Ricci tensor is defined as $R_\mn =R^\alpha {}_{\mu \alpha \nu}$.
\end{itemize}

%%%%%%%%%%%%%%%%%%%%%%%%%%%%%%%%%%%%%%%%%%%%%%%%%%%%%%%%%%%%%%%%%%%%%%%%%%%%%%%%%%%%
%%%%%%%%%%%%%%%%%%%%%%%%%%%%%%%%%%%%%%%%%%%%%%%%%%%%%%%%%%%%%%%%%%%%%%%%%%%%%%%%%%%%
%	f(R) gravity
%%%%%%%%%%%%%%%%%%%%%%%%%%%%%%%%%%%%%%%%%%%%%%%%%%%%%%%%%%%%%%%%%%%%%%%%%%%%%%%%%%%%
%%%%%%%%%%%%%%%%%%%%%%%%%%%%%%%%%%%%%%%%%%%%%%%%%%%%%%%%%%%%%%%%%%%%%%%%%%%%%%%%%%%%
\section{$f(R)$ gravity and the chameleon mechanism}\label{f(R)}
%%%%%%%%%%%%%%%%%%%%%%%%%%%%%%%%%%%%%%%%%%
%%%%%%%%%%%%%%%%%%%%%%%%%%%%%%%%%%%%%%%%%%
\subsection{$f(R)$ theories}\label{action}
We analyze metric $f(R)$ gravity, whose action is given by replacing the Ricci scalar $R$ in the Einstein-Hilbert action by a general function of $R$:
	\be
	S_{f(R)}=\fr{\mpl ^2}{2}\intel f(R)+S_{\rm m}[g_\mn ,\Psi ], \label{2.1}
	\ee
where $S_{\rm m}$ is the action for a matter field $\Psi$. Note that there exist other formulations of $f(R)$ gravity \cite{Sotiriou:2008rp}, such as the Palatini formalism \cite{Olmo:2011uz}, in which the metric and the connection are assumed to be independent variables.

Variation with respect to $g^\mn$ gives the equation of motion:
	\be
	R_\mn F(R)-\fr{1}{2}f(R)g_\mn =8\pi GT_\mn +\nabla _\mu \nabla _\nu F(R)-g_\mn \Box F(R), \label{2.2}
	\ee
where $F(R)\equiv df(R)/dR$ and $T_\mn$ is the stress-energy tensor for the matter field. From this equation one can see that $G_{\rm eff}\equiv G/F(R)$ acts as the gravitational constant in GR, which means $G_{\rm eff}$ depends on both position and time.

By taking the covariant derivative of Eq.~\eqref{2.2}, one can show that
	\be
	T^\mu {}_{\nu ;\mu}=0. \label{2.2a}
	\ee
So the equation of motion \eqref{2.2} is consistent with the conservation of energy-momentum of the matter field.

Now we introduce a scalar field $\phi$ by
	\be
	F(R)=e^{-2\beta \phi /\mpl} \label{2.3}
	\ee
where $\beta =1/\sqrt{6}$, and the Einstein-frame metric $\bg _\mn$ is
	\be
	\bg _\mn =e^{-2\beta \phi /\mpl}g_\mn . \label{2.4}
	\ee
Then the gravitational part of Eq.~\eqref{2.1} can be rewritten as
	\be
	S_{\rm gravity}=\int d^4x\sqrt{-\bg} \left[ \fr{\mpl ^2}{2}\bR -\fr{1}{2}\bn ^\la \phi \bn _\la \phi -V(\phi )\right] . \label{2.5}
	\ee
Here $V(\phi )$ is the potential of $\phi$:
	\be
	V(\phi )\equiv \fr{\mpl ^2}{2}\fr{RF(R)-f(R)}{F(R)^2}. \label{2.6}
	\ee
Note that the right-hand side can be thought of as a function of $\phi$ through Eq.~\eqref{2.3}.

In the Einstein frame, the equations of motion are as follows:
	\begin{align}
	\bR _\mn -\fr{1}{2}\bR \bg _\mn &=\kappa ^2\bn _\mu \phi \bn _\nu \phi -\kappa ^2 \bg _\mn \left( \fr{1}{2}\bn ^\la \phi \bn _\la \phi +V(\phi )\right) +\kappa ^2\bT _\mn , \label{2.7} \\
	\bar{\Box}\phi &=V'(\phi )-\fr{\beta}{\mpl}\bT . \label{2.8}
	\end{align}
Here a prime denotes a derivative with respect to $\phi$, $\bT _\mn$ is the stress-energy tensor and $\bT \equiv \bg ^\mn \bT _\mn$ is its trace. $\bT _\mn$ is related to the Jordan-frame quantity by
	\be
	\bT ^\mu {}_\nu =e^{4\beta \phi /\mpl}T^\mu {}_\nu . \label{2.9}
	\ee
Since $\phi$ is coupled to the matter field through $g_\mn$ in $S_{\rm m}$, the contribution of matter appears in Eq.~\eqref{2.8}. Due to this term, $\bT _\mn$ is not conserved in the Einstein frame. It is convenient to define a conserved quantity,
	\be
	\ti{T}^\mu {}_\nu =e^{-\beta \phi /\mpl}\bT ^\mu {}_\nu =e^{3\beta \phi /\mpl}T^\mu {}_\nu . \label{2.10}
	\ee
The corresponding densities in the two frames are related by $\ti{\rho}=e^{3\beta \phi /\mpl}\rho$. For $\phi \ll \mpl$, $\rho$ and $\ti{\rho}$ almost coincide with each other.

Now we have the problem of choosing a physical frame, {\it i.e.}, we have to match the Jordan- or Einstein-frame quantity with the observed one. In this paper we regard the Jordan frame as physical, where the masses of particles do not depend on their position or time. On the other hand,  it is difficult to solve Eq.~\eqref{2.2} directly since it is a fourth-order differential equation. Thus, we perform calculations in the Einstein frame where the equations of motion \eqref{2.7} and \eqref{2.8} are both secondorder.%%%%%%%%%%%%%%%%%%%%%%%%%%%%%%%%%%%%%%%%%%
%%%%%%%%%%%%%%%%%%%%%%%%%%%%%%%%%%%%%%%%%%
\subsection{Chameleon mechanism}\label{chameleon}
In the Newtonian limit, a test mass $M$ experiences a force $\vec{F}$ given by \cite{Khoury:2003rn,Waterhouse:2006wv}
	\be
	\fr{\vec{F}}{M}=-\fr{\beta}{\mpl}\vec{\nabla}\phi . \label{2.10a}
	\ee
This fifth force violates the weak equivalence principle and may result in disagreement with the local tests of GR.

In order for $f(R)$ theories to pass the local tests of gravity, the chameleon mechanism must work to suppress the fifth force compared to the Newtonian force \cite{Khoury:2003rn}. Since we are interested in the present era of accelerated expansion, we neglect radiation and assume the Universe is dominated by pressureless matter. The Klein-Gordon (KG) equation \eqref{2.8} becomes
	\be
	\bar{\Box}\phi =V'(\phi )+\fr{\beta}{\mpl}\ti{\rho}e^{\beta \phi /\mpl}=\Ve '(\phi ), \label{2.11}
	\ee
where  $\Ve (\phi )$ is the effective potential, which includes the contribution of the matter field:
	\be
	\Ve (\phi )\equiv V(\phi )+\ti{\rho}e^{\beta \phi /\mpl}. \label{2.12}
	\ee
If the functional form of $f(R)$ is designed to satisfy $V'(\phi )<0$, $V''(\phi )>0$ and $V'''(\phi )<0$, one can show the following.
\begin{itemize}
\item $\Ve (\phi )$ has a minimum.
\item The minimum $\pmin$ is a decreasing function of $\rho$.
\item The scalar mass at the minimum $m_\phi$ is an increasing function of $\rho$.
\end{itemize}
So in a dense region, $m_\phi$ takes a large value and the range of the fifth force becomes short. This is called the chameleon mechanism, and the corresponding scalar is called a chameleon field. By virtue of this mechanism, GR is restored on a local scale.

To see how the fifth force gets small concretely, let us consider a scalar field around a uniform spherical object with radius $R_c$, density $\rho _c$, and mass $M_c=4\pi R_c^3\rho _c/3$. The background spacetime has density $\rho _b$ and is assumed to be static. If the object is large enough, the value of $\phi$ inside the sphere is given by the minimum of $\Ve (\phi )$ whose shape is determined by $\rho _c$. This type of scalar configuration is known as the thin-shell solution \cite{Khoury:2003rn}.

The chameleon configuration is given by solving Eq.~\eqref{2.11} neglecting the time derivatives, {\it i.e.} the Poisson equation $\nabla ^2\phi =\Ve '(\phi )$. $\phi$ is assumed to sit at the potential minimum both well inside ($r<R_s<R_c$) and far from ($r\rightarrow \infty$) the object ($\phi _c$ and $\phi _b$, respectively), and the corresponding masses are denoted by $m_c$ and $m_b$\footnote{For simplicity, we assume $m_bR_c\ll 1$}. The solution is obtained as follows:
	\be
	\dph =\left\{ \begin{array}{ll}
	\displaystyle \dph _c,&r<R_s, \\
	\displaystyle \fr{\beta \rho _c}{6\mpl}\left( r^2+2\fr{R_s^3}{r}-3R_s^2\right) +\dph _c,&R_s<r<R_c, \\
	\displaystyle -\fr{\beta \rho _c}{\mpl}\eth \fr{R_c^3}{r}e^{-m_b(r-R_c)},&r>R_c.
	\end{array}\right. \label{2.13}
	\ee
Here  $\dph _c\equiv \phi _c-\phi _b$ and $\eth$ is a constant which parametrizes the thinness of the shell-like region $R_s<r<R_c$:
	\be
	\eth \equiv \fr{\mpl}{\beta}\fr{|\delta \phi _c|}{R_c^2\rho _c}\approx \fr{R_c-R_s}{R_c}. \label{2.14}
	\ee
This means that, for an object to have a thin-shell solution, $\eth <1$ is needed. We call this inequality the thin-shell constraint.

Note that $\eth$ can be rewritten as
	\be
	\eth =\fr{\beta |\delta \phi _c|/\mpl}{GM_c/R_c}, \label{2.15}
	\ee
which is considered as the ratio between the fifth force and the Newtonian potential. Therefore the thin-shell constraint implies that the fifth force is smaller than the Newtonian force. From the Solar System tests of the weak equivalence principle using the free-fall acceleration of the Earth and the Moon toward the Sun, it is known that $\eth$ for the Earth ($\ep _{{\rm th},\oplus}$) must be smaller than $2.2\times 10^{-6}$ \cite{DeFelice:2010aj}.

%%%%%%%%%%%%%%%%%%%%%%%%%%%%%%%%%%%%%%%%%%%%%%%%%%%%%%%%%%%%%%%%%%%%%%%%%%%%%%%%%%%%
%%%%%%%%%%%%%%%%%%%%%%%%%%%%%%%%%%%%%%%%%%%%%%%%%%%%%%%%%%%%%%%%%%%%%%%%%%%%%%%%%%%%
%	EoS for dark energy
%%%%%%%%%%%%%%%%%%%%%%%%%%%%%%%%%%%%%%%%%%%%%%%%%%%%%%%%%%%%%%%%%%%%%%%%%%%%%%%%%%%%
%%%%%%%%%%%%%%%%%%%%%%%%%%%%%%%%%%%%%%%%%%%%%%%%%%%%%%%%%%%%%%%%%%%%%%%%%%%%%%%%%%%%
\section{Equation of state of dark energy}\label{EoS}
%%%%%%%%%%%%%%%%%%%%%%%%%%%%%%%%%%%%%%%%%%
%%%%%%%%%%%%%%%%%%%%%%%%%%%%%%%%%%%%%%%%%%
\subsection{$f(R)$ cosmology}\label{cosmology}
From now on we impose the flat Friedmann-Lema$\hat{\i}$tre-Robertson-Walker metric in the Jordan frame:
	\be
	ds^2=-dt^2+a(t)^2\left[ dr^2+r^2(d\theta ^2+\sin ^2\theta ~d\varphi ^2)\right] . \label{3.1}
	\ee
Transformed into the Einstein frame, Eq.~\eqref{2.7} becomes
	\begin{align}
	&H^2-\fr{2\beta}{\mpl}H\dot{\phi}=\fr{1}{3\mpl ^2}\left[ V(\phi )F+\fr{\rhom}{F}\right] , \label{3.2a} \\
	&\fr{2\ddot{a}}{a}+H^2-\fr{2\beta}{\mpl}\left( \ddot{\phi}+2H\dot{\phi}\right) +\fr{2}{3\mpl ^2}\dot{\phi}^2=\fr{1}{\mpl ^2}V(\phi )F. \label{3.2b}
	\end{align}
To interpret these equations in terms of GR, we reassemble Eqs.~\eqref{3.2a} and \eqref{3.2b} as follows:
	\begin{align}
	H^2&=\fr{\rhom +\rhode}{3\mpl ^2F_0}\equiv H^2(\Omega _{\rm m}+\Omega _{\rm DE}), \label{3.3a} \\
	\fr{2\ddot{a}}{a}+H^2&=-\fr{P_{\rm DE}}{\mpl ^2F_0}. \label{3.3c}
	\end{align}
where $\Omega _{\rm m}, \Omega _{\rm DE}$ are the effective density parameters of matter and dark energy, respectively,
	\be
	\Omega _{\rm m}\equiv \fr{\rhom}{3H^2\mpl ^2F_0},~~~\Omega _{\rm DE}\equiv \fr{\rhode}{3H^2\mpl ^2F_0}. \label{3.3b}
	\ee
Equations \eqref{3.3a} and \eqref{3.3c} mean that we regard the present value of the effective gravitational constant $G_{\rm eff}$ as the gravitational constant of GR, and any deviation from GR is considered as dark energy:
	\begin{align}
	\fr{\rhode}{\mpl ^2F_0}&\equiv \fr{F}{\mpl ^2}V(\phi )+\fr{6\beta}{\mpl}H\dot{\phi}+\fr{\rhom}{\mpl ^2F_0}\left( \fr{F_0}{F}-1\right) , \label{3.4} \\
	\fr{P_{\rm DE}}{\mpl ^2F_0}&\equiv -\fr{F}{\mpl ^2}V(\phi )+\fr{2}{3\mpl ^2}\dot{\phi}^2-\fr{2\beta}{\mpl}(\ddot{\phi}+2H\dot{\phi}). \label{3.4a}
	\end{align}
Together with the effective equation of state of dark energy
	\be
	\fr{P_{\rm DE}}{\rhode}=w, \label{3.5a}
	\ee
we get
	\be
	(1+w)\Omega _{\rm DE}=\fr{\rhode +P_{\rm DE}}{\rho _{\rm cr}}=\fr{2\beta}{3\mpl}\left( -\fr{\ddot{\phi}}{H^2}+\fr{\dot{\phi}}{H}\right) +\fr{2}{9\mpl ^2}\fr{\dot{\phi}^2}{H^2}+\Omega _{\rm m} \left( \fr{F_0}{F}-1\right) . \label{3.6}
	\ee
Since time derivatives are of order $H$, this expression can be estimated as
	\be
	|(1+w)\Omega _{\rm DE}|\sim \mO \left( \fr{\beta}{\mpl}\Delta \phi \right) , \label{3.7}
	\ee
where $\Delta \phi$ is the variation of $\phi$ in the last Hubble time.

Now let us apply the thin-shell constraint to a cosmological situation, following the steps of Ref.~\cite{Brax:2008hh}. Here the following assumptions are made.
	\begin{enumerate}
	\item The Universe is approximately homogeneous when coarse grained over scales larger than some $L_{\rm hom}\ll H^{-1}$.
	\item Gravity is weak, {\it i.e.} $\Pn \ll 1$ and $v^iv^i\ll 1$.
	\item The chameleon mechanism works similarly as in the static case.
	\end{enumerate}
Since $\rho _c>\rho _b(t)>\rho _b(t_0)$ for a past time $t$ at which $z\gs 1$, the following relation holds by virtue of the chameleon mechanism:
	\be
	\phi _c<\phi _b(t)<\phi _b(t_0). \label{3.8a}
	\ee
If one identifies $\phi _b(t_0)-\phi _b(t)$ as $\Delta \phi$ in Eq.~\eqref{3.7}, it follows that
	\be
	|(1+w)\Omega _{\rm DE}|<\fr{\beta}{\mpl}(\phi _b(t_0)-\phi _c). \label{3.8b}
	\ee
On the other hand, the thin-shell constraint says that
	\be
	\fr{\beta}{\mpl}(\phi _b(t_0)-\phi _c)<\Phi _{\rm N} \label{3.8c}
	\ee
where $t_0$ is the present time. Combining Eqs.~\eqref{3.8b} and \eqref{3.8c}, one gets
	\be
	|(1+w)\Omega _{\rm DE}|<\Pn , \label{3.8}
	\ee
where $\Pn$ is the Newtonian potential of the celestial object under consideration. Since $\Pn$ is $10^{-6}-10^{-5}$ for large clusters and superclusters \cite{Lombriser:2012nn}, the following constraint is obtained:
	\be
	|(1+w)\Omega _{\rm DE}|<10^{-4}. \label{3.9}
	\ee
This can be thought of as a constraint on $w$ itself, since we already know $\Omega _{\rm DE}\approx 0.7$ \cite{Planck:2015xua}. The constraint \eqref{3.9} would imply that $f(R)$ gravity is indistinguishable from the cosmological constant model, as far as the evolution of the homogeneous background is concerned.

However, this constraint is physically unacceptable because a background cosmological quantity like $w$ should not be determined only by local information. This unphysical result was derived due to the assumption (iii). Actually, for models which predict $m_\phi \sim \mO (H)$, $\phi$ does not sit at the minimum of $\Ve (\phi )$ and Eq.~\eqref{3.8a} no longer holds. Furthermore, the original exterior solution, which was derived neglecting the time-derivative terms, does not satisfy the KG equation \eqref{2.11}. Therefore one must solve the correct field equation for such models, which will remedy the constraint of Ref.~\cite{Brax:2008hh}.

As a specific example, let us adopt Starobinsky's model \cite{Starobinsky:2007hu,Motohashi:2010tb}:
	\be
	f(R)=R+\la R_s\left[ \left( 1+\left( \fr{R}{R_s}\right) ^2\right) ^{-n}-1\right] \label{3.13}
	\ee
with $n, \lambda >0$ and $R_s$ is of the order of the observed cosmological constant. The parameter space can be constrained by both background-level and perturbative-level arguments \cite{Motohashi:2010tb}. At the background level, the asymptotic de Sitter solution should be stable. The smallest values of $\lambda$ which satisfy the stability condition are 0.95, 0.73, and 0.61 for $n=2$, 3, and 4, respectively. At the perturbative level, assuming that the ratio of the linear density perturbation $\delta _{f(R)}/\delta _{\Lambda {\rm CDM}}$ is small, the constraint resembles Fig. 5 of Ref.~\cite{Motohashi:2010tb}. In order to keep the ratio smaller than 10\% at $k=0.174h~{\rm Mpc}^{-1}$ (which is the wave number corresponding to $\sigma _8$ normalization), $\lambda$ should be larger than 8.2, 3.0, and 1.9 for $n=2$, 3, and 4 respectively. One can calculate $\ep _{{\rm th},\oplus}$ in this model and it turns out to be small enough to satisfy the Solar System tests of the weak equivalence principle. For example, for $n=2$ and $\lambda =1$, $\ep _{{\rm th},\oplus} \approx 10^{-17}$.

It is known that these models predict $|1+w|\sim \mO (0.1)$ and $m_\phi \sim \mO (H)$. So if the solution for the chameleon field with a thin shell is obtained in this framework, it can be thought of as a counterexample of the previous work \cite{Brax:2008hh}.
%%%%%%%%%%%%%%%%%%%%%%%%%%%%%%%%%%%%%%%%%%
%%%%%%%%%%%%%%%%%%%%%%%%%%%%%%%%%%%%%%%%%%
\subsection{Chameleon configuration in the accelerating Universe}
From now on we calculate a configuration for the chameleon field around a spherical object in a case where the deviation of $w$ from $-1$ is fairly large. The following assumptions are made.
	\begin{itemize}
	\item The background spacetime evolves as in the $w$CDM model (CDM+dark energy with constant $w$).
	\item The object is decoupled from cosmic expansion and its radius $R_c$ is constant in physical coordinates, {\it i.e.}, its surface is at $ar=R_c$.
	\item The interior ($ar<R_c$) solution $\dphi$ is given by Eq.~\eqref{2.13}, in which $r$ is replaced by $ar$.
	\item The exterior ($ar>R_c$) solution $\dpho$ is smoothly connected to $\dphi$.
	\item $\dph$ approaches zero at infinity.
	\end{itemize}
Here $\dph$ denotes the deviation of $\phi$ from the cosmological background value $\phi _b(t)$, which does not necessarily correspond to the minimum of $\Ve (\phi )$.

First we consider the de Sitter ($w=-1$) case, for which we can obtain the solution analytically, followed by the $w\ne -1$ case. In the following we use conformal time $\eta$ as the time variable.
%%%%%%%%%%%%%%%%%%%%%%%%%%%%%%%%%%%%%%%%%%
\subsubsection{$w=-1$ case}
The KG equation outside of the object becomes
	\be
	-\dph ''-2\mH \dph '+\nabla ^2\dph -m_b^2a^2\dph =0, \label{3.14}
	\ee
where $\mH \equiv a'/a$, a prime denotes a derivative with respect to $\eta$, and $m_b$ is the background mass of $\phi$. Since $\phi =-(\mpl /2\beta )\ln f'(R)={\rm const}$ in de Sitter spacetime, we neglect the $\eta$ dependence of the background $\phi$ and $m_b$. Using $a =-(H\eta )^{-1}$ for $w=-1$, Eq.~\eqref{3.14} can be written as follows:
	\be
	-\dph ''+\fr{2}{\eta}\dph '+\nabla ^2\dph -\fr{m_b^2}{H^2\eta ^2}\dph =0. \label{3.15}
	\ee
The boundary conditions are
	\begin{itemize}
	\item $\dpho$=$\dphi$ at $ar=R_c$,
	\item $\pa _r\dpho$=$\pa _r\dphi$ at $ar=R_c$,
	\item $\dpho \rightarrow 0$ as $ar\rightarrow \infty$.
	\end{itemize}

\begin{figure}[!ht]
\begin{center}
\includegraphics[width=10cm]{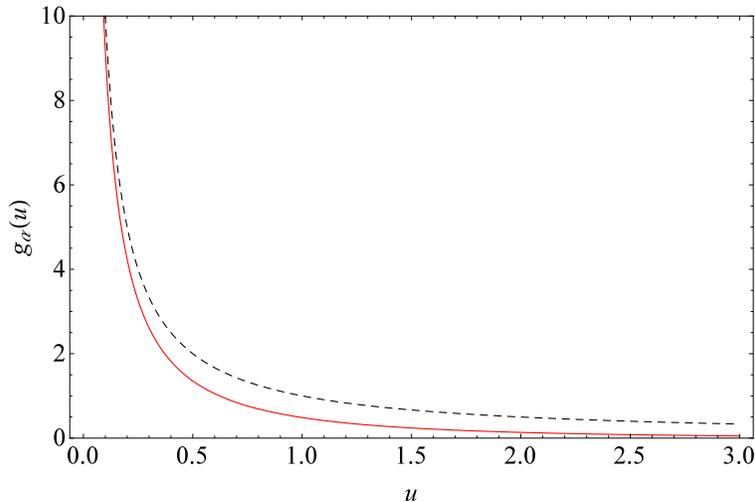}
\caption{The exterior solution $g_\alpha (u)$ for $\alpha =1$. The dashed line is a plot of $1/u$.}
\label{fig3.1}
\end{center}
\end{figure}

Let us describe the solution $\dph$ in terms of conformal time and the physical distance instead of the comoving distance, {\it i.e.}, in the form
	\be
	\dph (\eta ,r)=\vap (\eta ,u), \label{3.16}
	\ee
where $u$ denotes the physical distance normalized by $H^{-1}$:
	\be
	u\equiv Har=\mH r=-\fr{r}{\eta}. \label{3.17}
	\ee
Then Eq.~\eqref{3.15} becomes
	\be
	-\pa _\eta ^2 \vap -\fr{2u}{\eta}\pa _\eta \pa _u \vap +\fr{2}{\eta}\pa _\eta \vap -\fr{1}{\eta ^2}\left[ (u^2-1)\pa _u^2\vap +\left( 4u-\fr{2}{u}\right) \pa _u\vap +\left( \fr{m_b}{H}\right) ^2\vap \right] =0. \label{3.17a}
	\ee
The expression inside the square brackets of Eq.~\eqref{3.17a} is written only with $u$. Also note that the boundary conditions for $\vap$ are
	\begin{itemize}
	\item $\vap _{\rm out}$=$\vap _{\rm in}$ at $u=HR_c$,
	\item $\pa _u\vap _{\rm out}$=$\pa _u\vap _{\rm in}$ at $u=HR_c$,
	\item $\vap _{\rm out} \rightarrow 0$ as $u\rightarrow \infty$,
	\end{itemize}
and are expressed only with $u$. So we investigate a solution that is independent of $\eta$, {\it i.e.}, $\vap =\vap (u)$. In such a case, Eq.~\eqref{3.17a} can be rewritten as the following ordinary differential equation:
	\be
	\fr{d^2\vap}{du^2}+\fr{4u^2-2}{u(u^2-1)}\fr{d\vap}{du}+\fr{(m_b/H)^2}{u^2-1}\vap =0. \label{3.18}
	\ee
The independent solutions are obtained in terms of the hypergeometric function $_2F_1$ as follows:
	\begin{align}
	\vap _\alpha ^{(1)}(u)&={}_2F_1\left( \fr{3+2i\alpha}{4},\fr{3-2i\alpha}{4};\fr{3}{2};u^2\right) , \label{3.19a} \\
	\vap _\alpha ^{(2)}(u)&=\fr{1}{p}{}_2F_1\left( \fr{1+2i\alpha}{4},\fr{1-2i\alpha}{4};\fr{1}{2};u^2\right)  \label{3.19b}
	\end{align}
where
	\be
	\alpha \equiv \sqrt{\left( \fr{m_b}{H}\right) ^2-\fr{9}{4}}. \label{3.20}
	\ee
Both $\vap _\alpha ^{(1)}(u)$ and $\vap _\alpha ^{(2)}(u)$ diverge at $u=1$. Fortunately, we can construct a solution $g_\alpha (u)$ which is finite at $u=1$ by taking their linear combination:
	\be
	g_\alpha (u)\equiv \vap _\alpha ^{(2)}(u)-2\fr{\Gamma \left( \fr{3+2i\alpha}{4}\right) \Gamma \left( \fr{3-2i\alpha}{4}\right)}{\Gamma \left( \fr{1+2i\alpha}{4}\right) \Gamma \left( \fr{1-2i\alpha}{4}\right)}\vap _\alpha ^{(1)}(u). \label{3.21}
	\ee
Note that this function is defined for $u>0$ and takes a real value. The general form of $g_\alpha (u)$ is shown in Fig.~\ref{fig3.1}.

The exterior solution is proportional to $g_\alpha (\mH r)$, and the factor of proportionality is determined from the boundary conditions at the surface of the object $ar=R_c$, or $u=HR_c$. Using the fact that $g_\alpha (u)$ can be approximated\footnote{This approximation holds only if $HR_c\ll (m_b/H)\ll (HR_c)^{-1}$.} as $1/u$ for $u\approx HR_c\ll 1$, we get the following configuration for $\dph$:
	\be
	\dph =\left\{ \begin{array}{ll}
	\displaystyle \dph _c,&ar<R_s, \\
	\displaystyle \fr{\beta \rho _c}{6\mpl}\left[ (ar)^2+2\fr{R_s^3}{ar}-3R_s^2\right] +\dph _c,&R_s<ar<R_c, \\
	\displaystyle -\fr{\beta \rho _cR_c^2}{\mpl}\eth HR_cg_\alpha (\mH r),&ar>R_c.
	\end{array}\right. \label{3.22}
	\ee
Here $\eth$ is given by Eq.~\eqref{2.14}. Thus we managed to find the exterior solution that satisfies the boundary conditions.

%%%%%%%%%%%%%%%%%%%%%%%%%%%%%%%%%%%%%%%%%%
\subsubsection{$w\ne -1$ case}
Now we move on to the case where $w$ is a constant but not $-1$. We write $w=-1+\vae$ and consider up to first-order terms in $\vae$. In this case, $\mH$ is calculated as
	\be
	\mH =aH=-\left( 1+\fr{3}{2}\vae \right) \fr{1}{\eta} . \label{3.23}
	\ee
Therefore the equation of motion gets slightly modified from Eq.~\eqref{3.15}, and so does the solution. We assume the following form for the exterior solution:
	\be
	\dph =-\fr{\beta \rho _cR_c^2}{\mpl}\eth HR_c\left[ g_\alpha (\mH r)+\vae A(\eta ,r)\right] . \label{3.24}
	\ee
This expansion is valid if $A\ls \mO (1)$ at the horizon scale, since $g_\alpha (\mH r)\sim \mO (1)$ there (Fig.~\ref{fig3.1}). The equation for the perturbative part $A$ can be written as
	\be
	\vae [A''+2\mH A'-\nabla ^2A+m_b^2A]=\fr{\vae}{\eta ^2}\left[ -(2C_\phi -3)\mH rg_\alpha '(\mH r)-2C_\phi g_\alpha (\mH r)\right] , \label{3.25}
	\ee
where $C_\phi$ characterizes the rate of change of $\phi _b$,
	\be
	\fr{\phi _b'}{\phi _b}\equiv -\fr{C_\phi}{\eta}\vae . \label{3.26}
	\ee
Here $C_\phi \approx 3n$ for Starobinsky's model.

\begin{figure}[!ht]
\begin{minipage}{0.5\hsize}
\begin{center}
\includegraphics[width=10cm]{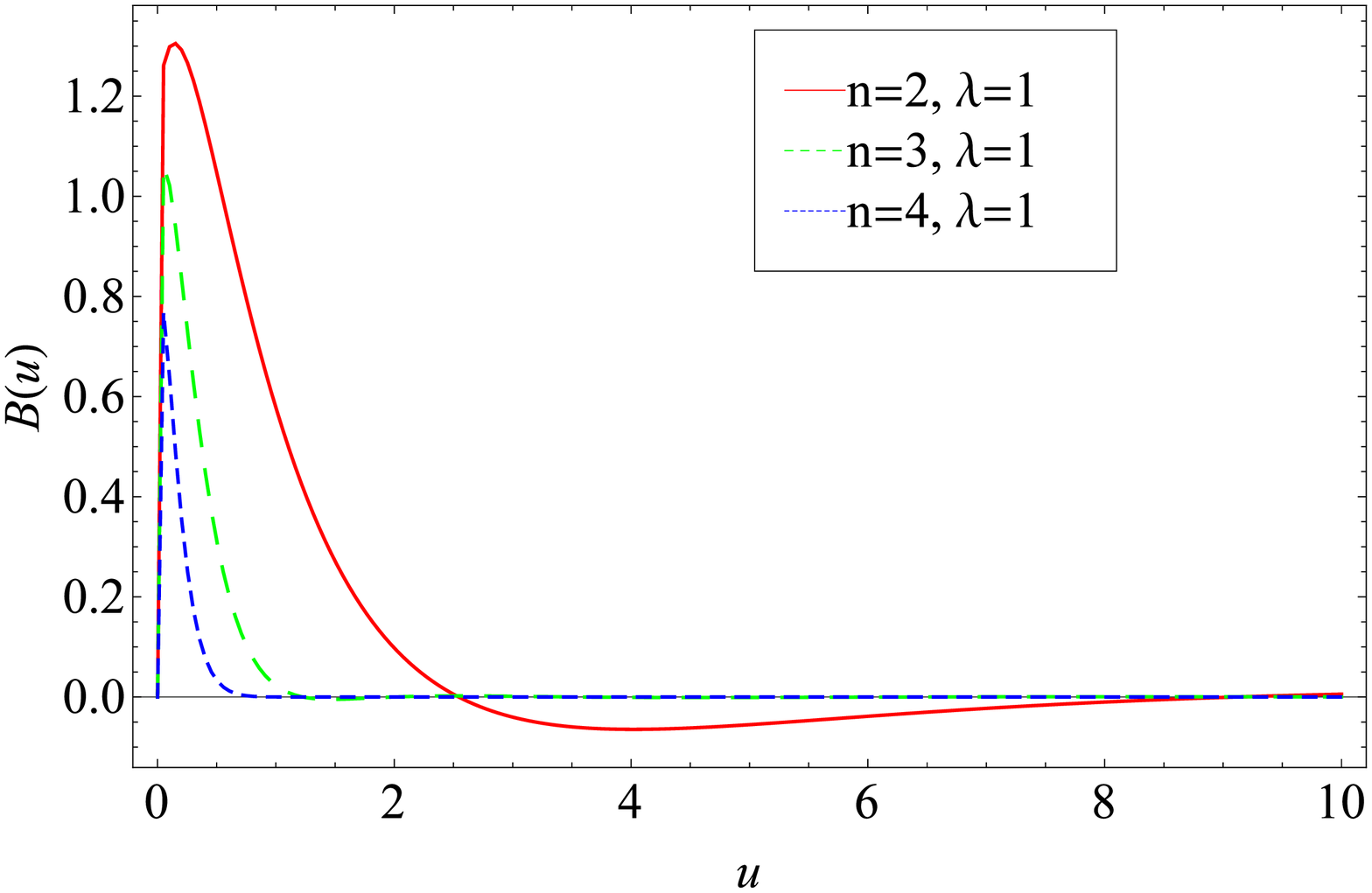}
\end{center}
\end{minipage}
\begin{minipage}{0.5\hsize}
\begin{center}
\includegraphics[width=10cm]{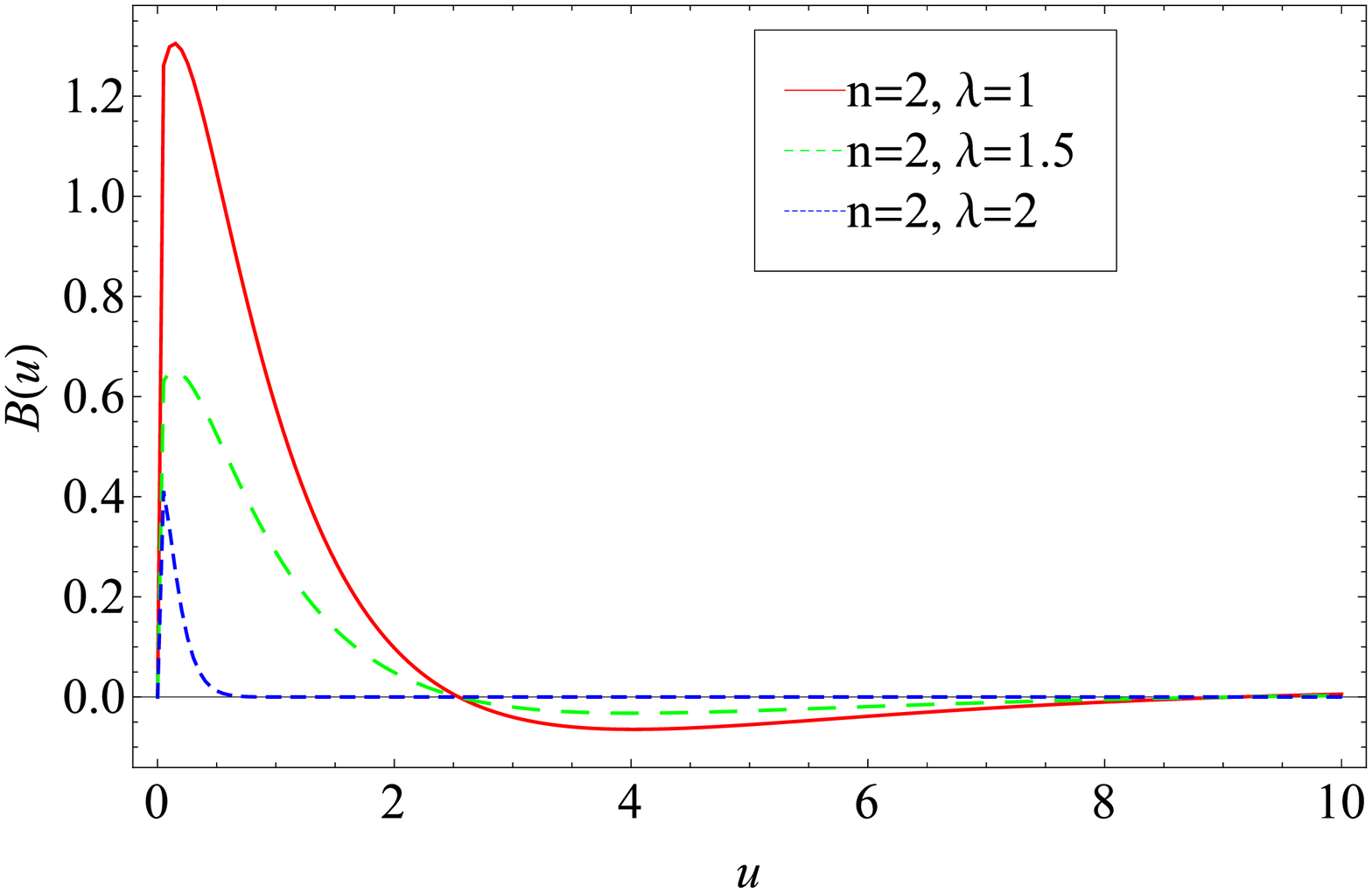}
\end{center}
\end{minipage}
\caption{The perturbative part $B(u)$ for various parameters of Starobinsky's model.}
\label{fig3.2}
\end{figure}

Again we try to find a solution in the form
	\be
	A(\eta ,r)=B(u),~~~u=-\fr{r}{\eta}. \label{3.27}
	\ee
Then Eq.~\eqref{3.25} is rewritten as an ordinary differential equation for $B(u)$:
	\be
	\fr{d^2B(u)}{du^2}+\fr{4u^2-2}{u(u^2-1)}\fr{dB(u)}{du}+\fr{(m_b/H)^2}{u^2-1}B(u)=j(u), \label{3.28}
	\ee
where the source term is given by
	\be
	j(u)\equiv -\fr{(2C_\phi -3)ug_\alpha '(u)+2C_\phi g_\alpha (u)}{u^2-1}. \label{3.29}
	\ee
Since we know the homogeneous solutions for Eq.~\eqref{3.28}, the inhomogeneous solutions can be obtained by the method of variation of parameters. We choose
	\bea
	B_1(u)&\equiv&\vap _\alpha ^{(1)}(u), \label{3.30a} \\
	B_2(u)&\equiv&g_\alpha (u) \label{3.30b}
	\eea
as a basis for the vector space spanned by the homogeneous solutions. Using these, the inhomogeneous solution for Eq.~\eqref{3.28} can be written as follows:
	\begin{align}
	B(u)&=C_1B_1(u)+C_2B_2(u)+B_s(u), \label{3.31} \\
	B_s(u)&\equiv -B_1(u)\int _0^udx\fr{B_2(x)}{W(x)}j(x)+B_2(u)\int _0^udx\fr{B_1(x)}{W(x)}j(x) \label{3.32}
	\end{align}
where $C_1, C_2$ are constants and $W(u)$ is the Wronskian of $B_1(u), B_2(u)$. The coefficient $C_1$ is fixed by assuming $B(u)$ does not diverge at $u=1$:
	\be
	C_1=\int _0^1dx\fr{B_2(x)}{W(x)}j(x). \label{3.33}
	\ee
$C_2$ is determined by requiring $B(u)=0$ at the surface of the object $u=HR_c$:
	\be
	C_2=-\fr{B_1(HR_c)}{B_2(HR_c)}\int _{HR_c}^1dx\fr{B_2(x)}{W(x)}j(x)-\int _0^{HR_c}dx\fr{B_1(x)}{W(x)}j(x). \label{3.34}
	\ee
Combining these, we get the solution for the perturbative part:
	\be
	B(u)=-B_1(u)\int _1^udx\fr{B_2(x)}{W(x)}j(x)+B_2(u)\left[ \int _{HR_c}^udx\fr{B_1(x)}{W(x)}j(x)-\fr{B_1(HR_c)}{B_2(HR_c)}\int _{HR_c}^1dx\fr{B_2(x)}{W(x)}j(x)\right] . \label{3.35}
	\ee
The form of $B(u)$ for various parameters of Starobinsky's model is shown in Fig.~\ref{fig3.2}. Here we set $HR_c=10^{-3}$, which corresponds to the ratio of the typical scale of a galaxy cluster to the present Hubble radius\footnote{The qualitative feature of the solution does not change drastically if $HR_c$ varies.}. Even in the $n=2, \lambda =1$ case, where the deviation from GR is the largest among the viable models, the magnitude of $B(u)$ is $\ls 1.3$. Thus we managed to obtain a consistent perturbative solution in the case of $w\ne -1$.

%%%%%%%%%%%%%%%%%%%%%%%%%%%%%%%%%%%%%%%%%%%%%%%%%%%%%%%%%%%%%%%%%%%%%%%%%%%%%%%%%%%%
%%%%%%%%%%%%%%%%%%%%%%%%%%%%%%%%%%%%%%%%%%%%%%%%%%%%%%%%%%%%%%%%%%%%%%%%%%%%%%%%%%%%
%	conclusion
%%%%%%%%%%%%%%%%%%%%%%%%%%%%%%%%%%%%%%%%%%%%%%%%%%%%%%%%%%%%%%%%%%%%%%%%%%%%%%%%%%%%
%%%%%%%%%%%%%%%%%%%%%%%%%%%%%%%%%%%%%%%%%%%%%%%%%%%%%%%%%%%%%%%%%%%%%%%%%%%%%%%%%%%%
\section{Conclusion}\label{conclusion}
Transformed into the Einstein frame, $f(R)$ gravity predicts a fifth force due to a scalar which is nonminimally coupled to matter. The fifth force is small if an object has a thin-shell configuration of the scalar field, and this enables $f(R)$ theories to pass local tests of gravity. If the thin-shell constraint is naively applied to cosmological scales, it leads to an extremely small $|1+w|$. This result was derived by solving the Poisson equation for the scalar field, which is inappropriate because the evolution of the scalar field is determined by the Klein-Gordon equation. By solving the correct field equation, we have shown that a consistent solution exists even if $w$ deviates appreciably from$-1$ as long as other viability conditions are satisfied.

\vskip 2cm
\noindent
{\large\bf Acknowledgments}

This work was partially supported by JSPS Grant-in-Aid for Scientific Research 23340058 (J.Y.).

\end{document}